\documentstyle[12pt,aaspp4]{article}
\begin{document}

\newcommand{\up}[1]{\ifmmode^{\rm #1}\else$^{\rm #1}$\fi}
\newcommand{\zdot}{\makebox[0pt][l]{.}}
\newcommand{\upd}{\up{d}}
\newcommand{\uph}{\up{h}}
\newcommand{\upm}{\up{m}}
\newcommand{\ups}{\up{s}}
\newcommand{\arcd}{\ifmmode^{\circ}\else$^{\circ}$\fi}
\newcommand{\arcm}{\ifmmode{'}\else$'$\fi}
\newcommand{\arcs}{\ifmmode{''}\else$''$\fi}

\title{The Araucaria Project. The Distance to the Local Group Galaxy
NGC 6822 from Cepheid Variables discovered in a Wide-Field Imaging Survey
\footnote{Based on  observations obtained with the 1.3~m
telescope at the Las Campanas Observatory.
}
}

\author{Grzegorz Pietrzy{\'n}ski}
\affil{Universidad de Concepci{\'o}n, Departamento de Fisica, Astronomy
Group,
Casilla 160-C,
Concepci{\'o}n, Chile}
\affil{Warsaw University Observatory, Al. Ujazdowskie 4,00-478, Warsaw,
Poland}
\authoremail{pietrzyn@hubble.cfm.udec.cl}
\author{Wolfgang Gieren}
\affil{Universidad de Concepci{\'o}n, Departamento de Fisica, Astronomy Group, 
Casilla 160-C, 
Concepci{\'o}n, Chile}
\authoremail{wgieren@astro-udec.cl}
\author{Andrzej Udalski}
\affil{Warsaw University Observatory, Aleje Ujazdowskie 4, PL-00-478,
Warsaw,Poland}
\authoremail{udalski@astrouw.edu.pl}
\author{Fabio Bresolin}
\affil{Institute for Astronomy, University of Hawaii at Manoa, 2680 Woodlawn 
Drive, 
Honolulu HI 96822, USA}
\authoremail{bresolin@ifa.hawaii.edu}
\author{Rolf-Peter Kudritzki}
\affil{Institute for Astronomy, University of Hawaii at Manoa, 2680 Woodlawn 
Drive,
Honolulu HI 96822, USA}
\authoremail{kud@ifa.hawaii.edu}
\author{Igor Soszy{\'n}ski}
\affil{Universidad de Concepci{\'o}n, Departamento de Fisica, Astronomy Group, 
Casilla 160-C,
Concepci{\'o}n, Chile}
\affil{Warsaw University Observatory, Aleje Ujazdowskie 4, PL-00-478, Warsaw, 
Poland}
\authoremail{soszynsk@astrouw.edu.pl}
\author{Micha{\l} Szyma{\'n}ski}
\affil{Warsaw University Observatory, Aleje Ujazdowskie 4, PL-00-478,
Warsaw, Poland}
\authoremail{msz@astrouw.edu.pl}
\author{Marcin Kubiak}
\affil{Warsaw University Observatory, Aleje Ujazdowskie 4, PL-00-478,
Warsaw, Poland}
\authoremail{mk@astrouw.edu.pl}

\begin{abstract}
We have obtained mosaic images of NGC 6822 in V and I bands on 77 nights. From these data,
we have conducted an extensive search for Cepheid variables over the entire field
of the galaxy, and we have found 116 such variables with periods ranging from
1.7 to 124 days. We used the long-period ($>$ 5.6 days) Cepheids to establish the
period-luminosity relations in V, I and in the reddening-independent Wesenheit
index, which are all very tightly defined. Fitting the OGLE LMC slopes in the various
bands to our data, we have derived distance values for NGC 6822 in V, I and ${\rm W}_{\rm I}$
which agree very well among themselves. Our adopted best distance value from the
reddening-free Wesenheit index is 23.34 $\pm$ 0.04 (statistical) $\pm$ 0.05 (systematic) mag.
This value agrees within the combined 1 sigma uncertainties with a previous distance
value derived for NGC 6822 by McAlary et al. from near-IR photometry of 9 Cepheids,
but our new value is significantly more accurate. We compare the slopes of the Cepheid
PL relation in V and I as determined in the five best-observed nearby galaxies, which
span a metallicity range from -1.0 to -0.3 dex, and find the data consistent with
no metallicity dependence of the PL relation slope in this range. Comparing the magnitudes
of 10-day Cepheids with the I-band magnitudes of the TRGB in the same set of galaxies,
there is no evidence either for a significant variation of the period-luminosity
zero points in V and I. The available data limit such a zero point variation to less
than 0.03 mag, in the considered low-metallicity regime.
\end{abstract}

\keywords{distance scale - galaxies: distances and redshifts - galaxies:
individual: NGC 6822 - galaxies: stellar content - stars: Cepheids}

\section{Introduction}
Nearby galaxies offer a unique laboratory to study in much detail how stellar
distance indicators depend on 
environmental properties like age and metallicity. Calibrating these dependences
accurately for the most important stellar candles is a fundamental requirement
in order to improve the distance scale. The enormous spread in the distance
to the LMC derived from different stellar objects and techniques (Benedict et al. 2002)
illustrates the magnitude of the problem, showing that environmental dependences
of stellar candles like Cepheids, red clump giants and RR Lyrae stars are not
yet under control to the point that the systematic uncertainties in these methods
are beat down below a few percent. This affects directly the calibration of the
distance scale, and of cosmological parameters linked to it one way or the other.
In our ongoing Araucaria Project (see Gieren et al. 2001; also
{\it http://ifa.hawaii.edu/\~{}bresolin/Araucaria/index.html}), we are
addressing the problem by observing large samples of different stellar distance
indicators in a sample of nearby galaxies in the Local and Sculptor Groups which
cover a broad range in their environmental parameters. Systematic comparison of
distance results obtained from the different candles should allow us to recognize
and calibrate the dependences on metallicity and age, and possibly other environmental
parameters, of the methods of distance determination we are scrutinizing, and to do this
with a significantly higher level of accuracy than what has been achieved
by other researchers in the literature. First
results, for instance on red clump stars, have already been obtained and discussed
(Pietrzynski \& Gieren 2002; Pietrzynski et al. 2003).

A first step in the exploration of distance indicators in the target galaxies
of the Araucaria Project consists in identifying sizeable samples of these objects
in each of them. With the advent of 
the new wide-field mosaic optical detectors, this can be done much more efficiently
than in those times when single CCD detectors, or even photographic surveys had
to be used to acquire the necessary data, which were then mostly restricted to small
parts of these galaxies. For all of our target galaxies
it is now possible to obtain high quality CCD photometry for the whole stellar content
in a reasonably short time, and therefore significantly
improve on results of previous
investigations which were derived from the study of small parts of these galaxies only.
Our first extensive wide-field survey was
performed in the Sculptor spiral galaxy NGC 300, and resulted in the 
discovery of 117 Cepheids and 12 Cepheid candidates (Pietrzynski et al. 
2002) which were recently used to determine the distance to this galaxy from
its population of long-period Cepheids (Gieren et al. 2004). We also discovered a 
considerable number of spectroscopically confirmed blue supergiants in this galaxy
from our wide-field survey (Bresolin et al. 2002), which are very useful objects
for distance determination out to several Megaparsecs from the Wind Momentum-
Luminosity Relationship (Kudritzki et al. 1999), or from the Flux-Weighted 
Gravity-Luminosity Relationship (Kudritzki et al. 2003; Urbaneja et al. 2004) 
they have been demonstrated to obey from our NGC 300 data.

As the next target in our survey for distance indicators, we selected the Local Group 
irregular galaxy NGC 6822.
This galaxy is a key object in the Araucaria Project since it contains all
the distance indicator classes we study in large numbers, and since it is one
of the nearest, and therefore best-resolved galaxies in our target list. Furthermore,
the stellar objects are obviously relatively bright, as compared to the more
distant galaxies in our target list, facilitating accurate photometry and spectroscopy.
NGC 6822 was already subjected to a search for variable stars a long time ago using
photographic plates (Hubble 1925, Kayser 1967),  which led to 
discovery of 21 Cepheids. Subsamples of these Cepheids were later observed using 
CCD detectors in both, the near-IR H band (McAlary et al. 1983 - 9 Cepheids)
and optical BVRI bands (Gallart et al. 1996 - 6 Cepheids) in order 
to improve the distance determination to NGC 6822. Recently, Antonello
et al. (2002) monitored a small field  of 3.8 x 3.8 arcmin, located 
at the centre of NGC 6822, in white light (no filter used), finding 15 new 
population I Cepheids in this region. 

Encouraged by this recent result, we performed a wide-field mosaic imaging survey
in NGC 6822 which covers the entire galaxy,
 with the main goal to discover a significant number of additional
Cepheid variables, particularly of long-period Cepheid variables, and other
objects we will be studying in our project in forthcoming papers. Using an
improved Cepheid database (both in quantity and in quality), our goal was 
to  significantly improve previous distance determinations to this galaxy using
the reddening-insensitive Wesenheit magnitude-period relation for Cepheids
which we successfully applied in our recent work on NGC 300 (Gieren et al. 2004).
 In this paper, which is focused on a
distance determination from Cepheid variables, we only present the Cepheids we discovered
in our mosaic survey. Other types of variables we have also discovered will be presented
 and discussed in a forthcoming paper, as will be the abundant blue supergiant population 
 in NGC 6822 we have discovered from our data.
 
 In section 2, we present and discuss the observations, calibrations and
 data reduction procedures. In section 3, we present the catalog of Cepheid variables
 in NGC 6822 obtained from our data, and in section 4 we use these data to
 determine the distance to NGC 6822 from the Cepheid PL relation. In section 5,
 we discuss our result, and its uncertainty, and in section 6 we summarize
 the main conclusions of this paper.

\section{Observations,  Reductions and Calibrations}
All the  data presented in this paper were collected with the Warsaw 1.3-m 
telescope at Las Campanas Observatory. The telescope was equipped with 
a mosaic 8k $\times$ 8k detector, with a field of view of about 35 $\times$ 35 
arcmin and a scale of about 0.25 arcsec/pix. For more  instrumental
details on this camera, the reader is referred to the OGLE  website:
{\it http://ogle.astrouw.edu.pl/\~{}ogle/index.html}.
V and I band  images of NGC 6822 were secured on 77 different nights.
The exposure time was set to 600 seconds in both filters.
Most of the observations were obtained in 2002 between July 3 and November 21.
Additional observations were made during three photometric nights in May 2003 
in order to accurately calibrate the data and improve periods of the
detected variable stars.

Preliminary reductions (i.e. debiasing and flatfielding)  were 
done with the IRAF\footnote{IRAF is distributed by the
National Optical Astronomy Observatories, which are operated by the
Association of Universities for Research in Astronomy, Inc., under cooperative
agreement with the NSF.} package. Then, the PSF photometry was obtained 
for all stars in the same manner as in Pietrzy{\'n}ski, Gieren and
Udalski (2002). Independently, the data were reduced with the OGLE III pipeline 
based on the image subtraction technique (Udalski 2003; Wo{\'z}niak 2000).  

In order to calibrate our photometry onto the standard system 
our target was monitored during three photometric nights
(2003 May 11, 17 and 28) together with 
some 25 standards from the Landolt fields spanning a wide range of
colors ( -0.14 $<$ V-I $<$ 1.43), and observed at very different
airmasses. Since in principle the transformation equations for each of the eight
chips may have different color coefficients and zero points, the selected
sample of standard stars was observed on each of the individuial chips, and
transformation coefficients  were derived independently for each chip, 
on each night. Comparison of the photometry obtained on the different nights
(see Fig. 1) revealed that the internal accuracy of zero points in both V and
I bands is better than 0.01 mag. 

To check in detail the possible variation of the zero points in V and I
over the mosaic, two fields in the SMC, which contain a huge number of uniformely
distributed  stars 
with very accurate photometry established by the OGLE team (Udalski 
et al. 1999a) were monitored on May 28 together with our target and 
standard fields, and were calibrated in the same manner as the NGC 6822 
observations. Typically, in the selected SMC fields about 200.000 stars 
were found in common with the OGLE II photometric maps on each of the eight
mosaic chips. Comparison of OGLE II and our photometry (see Fig. 2) 
shows that on average the difference in the zero points between 
these two photometric sets is smaller than 0.01 mag on each chip. 
The  scatter, being larger than expected from the accuracy of both data sources 
indicates that there is a small variation of the zero point (up to 
about 0.05 mag) over the mosaic field, in both filters. In order to account for this 
effect each chip was divided into 64  512 x 512 pixel ( about 2 x 2
arcmin) boxes, and the difference between our photometry and OGLE II
data was calculated for each box in both V and I bands. Adopting 
such a resolution we assure  very good statistics in each box, and 
therefore an accurate (rms smaller than 0.02 mag) determination of our 
"correction maps". 
After applying these correction maps in V and I to  
the SMC fields , the scatter of the difference between the  OGLE II and 
our new mosaic magnitudes for the stars in these fields was reduced  
down to 0.02-0.03 mag (see Fig. 3),
showing that our adopted correction process is indeed improving the accuracy
 of the photometry derived from the mosaic images. 
Finally, the NGC 6822 photometry was corrected with these same maps 
assuring that photometric zero point changes over the mosaic field are kept below 
0.03 mag in both, V and I filters. 

There are two sources of photometric data available in the
literature which allowed us to perform checks on our 
NGC 6822 photometry. Fig. 4 presents a comparison of the V band 
photometry of stars 
from one of the chips of our mosaic camera with photometry published by Bianchi et al.
(2001). As one can see, there is no difference between the zero points
of these two data sets. Unfortunately, Banchi et al. (2001) did not provide 
I band data. Fig. 5 shows the difference between  the photometry presented by Gallart 
et al. (1996)  and our results for one of the mosaic chips, in both V and I.
In the I band, the difference in the zero points amounts to about 0.03 mag. 
In the V band, there is practically no offset (e.g. the
difference is smaller than 0.005 mag) for stars fainter than 19 mag; however, 
for brighter stars there is an offset of about 0.03 mag (see upper
panel of Fig. 5). 
Since no such effect is present in the comparison of 
our data with the photometry obtained by Bianchi et al., and 19 mag in V just
corresponds to the saturation limit of long exposures secured by
Gallart et al. (1996), we suspect that this jump is associated with a
difference between the zero points of the photometry obtained from the long and short 
exposures of Gallart et al. (1996), which were merged into one
database. If we repeat the comparisons using data from other chips of our mosaic camera, 
the results look very similar.

Summarizing the previous discussion, we may conclude that all the external and internal checks 
we made indicate that the mean zero points of our data are accurate to 0.03 mag 
in both V and I, and that their possible variations over the mosaic 
fields are smaller than 0.03 mag.

\section{Cepheid Catalog}
All stars observed in NGC 6822 were searched for photometric variations with periods
between 0.2 and 100 days, which corresponds to the time baseline 
covered by our observations, using the AoV algorithm (Schwarzenberg-Czerny
1989). The accuracy of the derived periods is about $ 1 \times 10^{-4}
\times {\rm P}$. In order to distinguish Cepheids from other types 
of variable stars several criteria, similar to those used in 
Pietrzy{\'n}ski et al. (2002),  were adopted. All light curves 
showing the typical shape of Cepheid variables, were approximated 
by  Fourier series of order 4. Then we rejected all candidates with 
amplitudes in the V band smaller than 0.4 mag, which is approximately the lower limit
amplitude for normal classical Cepheids. For the stars remaining in our database,
mean V and I band magnitudes were derived by integrating their 
light curves previously converted onto an intensity scale, and converting the results 
back to the magnitude scale. Due to the very good quality and phase coverage
of our light curves, the statistical accuracy of the
derived mean V and I magnitudes was typically better
than 0.01 mag. In a final step of selecting the Cepheid sample in NGC 6822,
we deleted objects having 
mean magniudes which are widely different (by a full magnitude or more) from those of Cepheids of
similar periods, clearly standing off the PL relations presented in section 4
(these and other variables will be studied in a forthcoming paper). 

Finally, we ended up with a catalog of 116 Cepheids. 
Table 1 presents their description: identification, equatorial
coordinates, period, time of maximum light in V, mean V and I band 
magnitudes, their Wesenheit index defined as {$ {\rm W}_{\rm I}= {\rm I}
- 1.55 \times (<{\rm V}> - <{\rm I}>)$}, and a cross-identification.

Our list of Cepheids  contains 9 out of the 11 Cepheids catalogued by 
Hubble (1925). The stars designed as V10 and V11 by Hubble turned out 
not to be Cepheids. We also rediscovered all additional Cepheids discovered
by Kayser (1967), with the exception of her variable V30. This object
is located in a very dense region in NGC 6822 
and it was impossible to cross-identify it unambiguously with 
stars on our images. Six of the Cepheids reported by Antonello et al. (2002)
were located between the chips of our mosaic and therefore did not enter
our catalog. It should be stressed here that the gaps between 
the chips of the mosaic installed on the 1.3 m Warsaw telescope are 
very small (of order of 10 arcsec), so very few Cepheids are expected 
to be missed due to the gaps. We found that the star designated as 
V0771 and classified as a W Vir variable with a period of 16.74 days by Antonello et
al. seems to be another classical Cepheid. Our data yield a 
period of 4.6046 days for this star (cep060). Since its location on 
P-L diagrams fits very well with the locations of other Cepheids with 
similar periods, we decided to include this star into the present catalog.

Table 2 contains the individual new V and I observations for all the Cepheids 
in Table 1. The full Table 2 is available in electronic form. In Fig. 6, we
are displaying the light curves of several of the Cepheids in our catalog, which
are typical in quality for the light curves of other variables of similar periods.

\section{PL relations and distance determination}
For the purpose of studying the period-luminosity relations defined by the 
Cepheids in NGC 6822  in V, I and {${\rm W}_{\rm I}$}, 
we excluded all stars having periods 
larger than 100 days (1 object), and smaller than 5.6 days 
from the data given in Table 1. A period of 100 days is usually 
assumed as the  upper limit of validity of the PL relation for Cepheids
(for a closer discussion of this, see Gieren et al. 2004).
The PL relation for Cepheids with periods larger than 100 days becomes 
nearly flat, which can probably be explained by the nonlinear pulsation cycle 
characteristics (Aikawa and Antonello 2000). The short-period 
cutoff period of 5.6 days was adopted to avoid contamination 
of our sample by first overtone pulsators. As can be seen 
from the very impressive PL relations derived from 1333 Cepheids in 
the LMC, presented by the OGLE team (Udalski et al. 1999c, see their 
Figure 2), first overtone Cepheids start to appear on the PL diagram 
shortward of logP = 0.75, which corresponds to our adopted 
lower limit period of 5.6 days. In the adopted period range from 5.6 to 100 days, 
three additional stars, cep026, cep029 and cep057, were rejected from our sample 
because of their strongly deviating location on the P-L diagrams. These 3 objects 
have mean Wesenheit magnitudes more than 3 sigma different from the mean 
magnitudes of Cepheids with similar periods.  
Our final sample of Cepheids used for the analysis of the 
PL relations in NGC 6822  contains 42 objects. As a consequence of our selection process,
there is a very high probability that all of these variables are classical fundamental mode
pulsators. We also note that our final Cepheid sample contains a very considerable
number of long-period variables (21 with periods longer than 10 days), which are crucial
for an accurate distance determination with the period-luminosity relation.

Fits to a straight line to our data yield the following 
slopes for the PL relations: -2.66 $\pm$ 0.07, -2.92 $\pm$ 0.05 
and -3.32 $\pm$ 0.04 in V, I and ${\rm W_{\rm I}}$, respectively. These 
values agree  very well with the corresponding OGLE LMC slopes of -2.775, -2.977 
and -3.300 (Udalski 2000). In particular there is excellent agreement 
for the slope of the reddening-free ${\rm W_{\rm I}}$ index, crucial
for the distance determination. 
It is wortwhile noticing, at this point, that in recent work basically identical slopes for 
this relation were obtained for several galaxies which
span a large range in metallicities (e.g. 
SMC - Udalski et al. 1999b, IC 1613 - Udalski et al. 2001, NGC 300 
- Gieren et al. 2004), indicating that the Wesenheit (V-I) PL relation must be substantially
metallicity independent, making it a very powerful tool for
distance determination. A full analysis of the constancy of the PL slopes in different
photometric bands will be undertaken once we have obtained PL relations for all
the galaxies of the Araucaria Project.

Given the good agreement of the PL slopes in the different bands derived for NGC 6822
from our data, with the
LMC OGLE slopes, we chose to adopt the extremely well determined OGLE slopes to derive the 
distance to NGC 6822. This leads to the following 
equations:\\

V = -2.775 log P + (23.024 $\pm$ 0.033) \\

I = -2.977 log P + (22.118 $\pm$ 0.026) \\

${\rm W}_{\rm I}$ = -3.300 log P + (20.706  $\pm$ 0.021) \\

Adopting 18.50 mag as the distance modulus to the LMC, 
E(B-V) = 0.36 mag as the sum of foreground and intrinsic 
reddenings for NGC 6822 (McAlary et al. 1986), and the reddening law of 
Schlegel et al. (1998) ( ${\rm A}_{\rm V}$ = 3.24 E(B-V), ${\rm A}_{\rm V}$ = 1.96
E(B-V)) we obtain the following distance moduli for NGC 6822 in the
three different bands we study:\\

$(m-M)_{0}$ (${\rm W}_{\rm I}$) = 23.338 mag \\

$(m-M)_{0}$ (I) = 23.319 mag\\

$(m-M)_{0}$ (V) = 23.292 mag\\

As can be seen, the results from the three different bands 
are in excellent agreement. Possible sources  of
uncertainty affecting these values are discussed in the next section.
The very slightly shorter distances obtained in the V and I bands may 
be caused by a slight overestimation of the adopted reddening. Indeed,
reducing the assumed reddening by just 0.01 mag we would 
get practically the same distances (within 0.01 mag) 
from all three bands.

\section{Discussion}
Before arriving at our best value
for the distance to  NGC 6822, together with the associated statistical
and systematical errors (see the next section),  we will now discuss 
how the potential dependence of the PL relation on metallicity, 
blending with unresolved companion stars, and the uncertainty 
of the assumed reddening  may affect our 
result. Another contribution to the total error budget, the uncertainty 
of the zero point of our photometry, was already discussed in detail in 
section 2. We will not discuss here the probably largest contribution to 
the systematic error of the distance to NGC 6822, 
which is the current uncertainty of our adopted
LMC distance modulus. The reader is referred to the paper of 
Benedict et al. (2002) for a modern review of distance determinations
to the LMC. In our paper, we assume a true distance modulus 
to the LMC of 18.50 mag in order to place our distance of NGC 6822 onto 
the same scale as the Cepheid distances to some 30 galaxies 
 derived by the HST Key 
Proyect for the Extragalactic Distance Scale (Freedman et al. 2001),
and to be consistent with our recent distance determination
from Cepheid variables for NGC 300 (Gieren et al. 2004).

\subsection{Dependence of the PL relation on metallicity}
In order to investigate the possible dependence of the Cepheid
period-luminosity relation on the metallicity of the Cepheids,
from a purely empirical point of view,
we can compare the well-determined period-luminosity relation in NGC 6822
of this paper
to the PL results obtained for other galaxies of different metallicities.
According to the high spectral resolution study of Venn et al. (2001), 
the mean metallicity of the young stellar population in NGC 6822
is [Fe/H] = -0.5 dex, intermediate between LMC and SMC.
We select for comparison only those nearby galaxies for which large samples of
Cepheids have very well-measured light curves in the V and I bands, which
have allowed to establish the coefficients of their respective PL relations
with high accuracy. The list of these galaxies currently includes the LMC, SMC, 
IC 1613 and NGC 300 (and will be expanded in the course of our project). It is 
worthwhile noticing that it is reasonable to assume 
(see the next section) that for all these well-resolved galaxies 
blending does not significantly affect the measured Cepheid 
brightnesses, and therefore the slopes and zero points which were obtained 
from the analysis of the respective PL relations.

As we already mentioned in the previous section, the slopes of the
PL relation in the V, I and ${\rm W}_{\rm I}$ bands in NGC 6822
agree very well with the values of the slopes derived from the 
Cepheids in the LMC (Udalski et al. 1999b, Udalski 2000), 
SMC (Udalski et al. 1999b, Udalski 2000), IC 1613 (Udalski 
et al. 2001) and NGC 300 (Gieren et al. 2004). The differences
of the observed slopes in these five galaxies, in any of these bands,
do never deviate by more than about 1.5 sigma of the respective
determinations, which we interpret as a clear sign that from
the excellent Cepheid data available for this well-observed set
of galaxies, no significant difference in the slope of the
PL relation can be established. Recalling that these galaxies 
have mean metallicities of their Cepheid populations in the range
-1.0 $<$ [Fe/H] $<$ -0.3, our current work on NGC 6822 confirms and strengthens 
the earlier finding of Udalski et al. (2001)
that the slope of the Cepheid PL relation in the V and I bands seems to 
be universal in this metallicity range. As a note of caution, however,
we note that the situation may change 
for the higher-metallicity regime, where careful studies of Galactic
Cepheids of near-solar mean metallicity have provided some evidence 
for steeper PL slopes (e.g. Storm et al. 2004; Gieren et al. 1998). 
Studies to confirm this later finding must continue,
at an ever increasing level of accuracy.

To verify if and how the absolute zero point of the Cepheid PL relation depends
on the metal content we decided 
to use the I-band magnitude of the tip of the red giant branch (TRGB)
as the reference brightness for Cepheids. There are strong 
reasons to believe that such an approach leads to very robust 
results. I-band TRGB magnitudes can be derived very accurately from the excellent
number statistics provided by
wide-field surveys of nearby galaxies; the TRGB brightness is similar 
to the Cepheid brightness and, most importantly for our current discussion,
the TRGB I-band magnitude does not depend on metallicity for a wide range of metallicities 
(Ballazzini, Ferraro and Pacino 2001, and references therein). 
In particular, the comparison of the I-band magnitudes of Cepheids 
and the TRGB measured from the same data is purely differential 
and the errors of the photometric zero point  and the adopted reddening 
should cancel out, and therefore such a comparison can be done very precisely.

To determine the I-band magnitude of the TRGB in NGC 6822 from our data we selected 
the red giants from their positions on the CMD diagram. 
Fig. 10  presents the histogram of the brightness of the
selected red giants, constructed with 0.04 mag bins.
The location of the TRGB is clearly depicted in this diagram at 19.83 $\pm$ 0.03 mag.
This value agrees very well with the TRGB magnitude of 19.8 $\pm$ 0.1 mag previously
determined for NGC 6822 by Gallart et al. (1996). 
Finally, and following Udalski et al. (2001), we can derive the difference 
between the I-band magnitude of the TRGB and the I-band magnitude of Cepheids 
with a period of 10 days. The value for NGC 6822 from our data is 0.68 $\pm$ 0.05 mag.
As can be seen from Fig. 11, which compares this difference to the
corresponding values observed for the LMC, SMC and IC 1613 (see 
Udalski et al. (2001)), the difference value is constant to within 
0.03 mag for these galaxies, a result which strongly supports the idea
that, apart from the slopes in V and I, the absolute I-band zero point of the Cepheid PL relation 
is universal in the metallicity range from -1.0 and -0.3 dex. 

One may therefore conclude that adopting the OGLE LMC slopes 
and ZPs for the distance determination to NGC 6822 we do not introduce 
any significant systematic error, given the small metallicity difference between the
LMC (-0.3 dex) and NGC 6822 (-0.5 dex). 

\subsection{Blending}
As was shown by Mochejska et al. (2000, 2001), Cepheid 
magnitudes obtained from ground-based data with a typical 
resolution of 1.5 arcsec, may be systematically overestimated up 
to 0.07 mag (for Cepheids with periods longer than 10 days) due to crowding
 for relatively nearby galaxies 
like M31 and M33. Since NGC 6822 is located at a shorter distance of roughly a 
60 \% of the distance of M 33, and our images are of higher resolution
than those of M 33 used by Mochejska et al., it is reasonable to expect
that the effect of blending on the Cepheid magnitudes in NGC 6822 
reported in this paper will be smaller than 0.07 mag.
Indeed, Udalski 
et al. (2001) comparing ground-based and HST magnitudes of Cepheids 
in IC 1613, a galaxy of the same type and of similar distance as NGC 6822, concluded that 
blending does not affect the ground-based measurements by more than 
0.04 mag. Another argument in favour of a very small effect of crowding  
on our ground-based Cepheid magnitudes in NGC 6822 comes from a
simulation made by Stanek and Udalski (1999), who, using the 
OGLE data, analyzed how the magnitudes of the LMC Cepheids observed 
at a worse resolution (e.g. larger distance) would be affected by
crowding. Since the LMC is a galaxy of the same type as NGC 6822, 
we can use their results to predict how strong a blending effect 
one should expect for Cepheids in NGC 6822. As can be seen from 
Fig. 4 in Stanek and Udalski (1999), the increased blending would 
affect the LMC Cepheid magnitudes at an average level of 0.03 mag, 
only. 
In the light of the above discussion it is reasonable to assume 
that crowding should have an effect on our distance 
determination presented in this paper which does not exceed 0.04 mag,
in the sense that the true distance to NGC 6822 should not be
more than a maximum of 0.04 mag larger than our derived result.

\subsection{Reddening}
Unfortunately, NGC 6822 is located relatively close to the Galactic 
plane ( b = -18 deg), which results in a relatively large, 
and possibly differential reddening across the galaxy. 
However, the tightness of the
PL relations obtained in this paper for the Cepheids in NGC 6822
(see Figs 5-7) clearly suggests that the differential reddening to the Cepheids
is not strong. Furthermore, in our approach we have circumvented the 
problem of the lack of information about the individual reddenings of the Cepheids
by using the reddening-free Wesenheit index ${\rm W}_{\rm I}$. 
It is  remarkable that assuming a constant E(B-V) = 
0.36 mag as advocated by McAlary et al. (1983) we obtain fully consistent distances 
from the PL relations constructed in the V, I and ${\rm W}_{\rm I}$ bands, which again
indicates that differential reddening is not an important issue in NGC 6822, and that
the adopted mean reddening must be very nearly correct.

Finally, our best distance modulus for NGC 6822 of 23.34 mag agrees quite well with the
distance modulus of 23.47 $\pm$ 0.11 mag as derived by McAlary et al. (1983)
from observations of 9 Cepheids in the near-infrared H-band, where the influence of reddening 
is significantly reduced as compared 
to the optical bands used in our study. This supports our conclusion that uncertainties
due to reddening have a very small, and probably negligible influence on our
distance result.

\section{Conclusions}
Our best distance value for NGC 6822, which comes from the
reddening-free Wesenheit magnitudes is  23.34 $\pm$  0.04 (statistical) 
$\pm$ 0.05 (systematic). The quoted statistical error contains the
quadratically combined uncertainties of the OGLE and our fits to the PL
relations of the LMC and NGC 6822, respectively, while the systematic uncertainty
includes the quadratically combined 
uncertainties of the zero points of both photometric studies, and a contribution
due to blending. As remarked before, our distance value does {\it not} take
into account the systematic uncertainty of the adopted LMC distance of 18.50 mag.
Should future research provide an improved determination of the distance to the LMC,
our derived distance to NGC 6822 can be easily adjusted to such a change.
As extensively discussed in the previous section, our NGC 6822 distance value 
seems very little affected by any systematic error associated with a possible
dependence of the Cepheid PL relation on metallicity, and with reddening.
Taking this into account, plus the fact that our new
distance determination comes from a large sample of Cepheids (about 
6 times larger than the samples of Cepheids used for previous distance
determinations to NGC 6822) with very well established light curves
and periods, we conclude that the presented
result is the to-date most robust distance determination to this galaxy.

A comparison of the slopes of the V- and I-band PL relations obtained from the
currently best-observed galaxies (LMC, SMC, IC 1613, NGC 6822 and NGC 300) strongly
suggests that the slopes in both bands do not depend on metallicity, in the
range -1.0 $<$ [Fe/H] $<$ -0.3 defined by these galaxies. A comparison of the
I-band TRGB magnitudes with the magnitudes of Cepheids at a pulsation period
of 10 days in the same sample of galaxies does not provide any evidence for
a significant variation of the PL zero points in V and I with metallicity either. 
This finding evidently strengthens our confidence that Cepheid variables are
excellent standard candles, at least when they are used in the low-metallicity
regime.

\acknowledgments
We are grateful to Las Campanas 
Observatory, and to the CNTAC for providing the large amounts of 
telescope time which were necessary to complete this project.
GP and  WG  gratefully acknowledge 
financial support for this
work from the Chilean Center for Astrophysics FONDAP 15010003. 
WG also acknowledges support
from the Centrum fuer Internationale Migration und Entwicklung in
Frankfurt/Germany
who donated the Sun Ultra 60 workstation on which a substantial part of the data 
reduction and analysis for this project was carried out.   
Support from the Polish KBN grant No 2P03D02123 and BST grant for 
Warsaw University Observatory is also acknowledged.

\begin{deluxetable}{c c c c c c c c c}
\tablecaption{Cepheids in NGC 6822}
\tablehead{
\colhead{ID} & \colhead{RA} & \colhead{DEC}  & \colhead{P} & \colhead{ ${\rm
T}_{0}$} &
\colhead{$<V>$} & \colhead{$<I>$} & \colhead{$<W_{\rm I}>$} &
\colhead{Remarks} \\
 & \colhead{(J2000)} & \colhead{(J2000)}  &
\colhead{ [days]} & \colhead{-2450000} &
\colhead{[mag]} & \colhead{[mag]} & \colhead{[mag]} &
}

\startdata
cep001 & 19:45:02.01 & -14:47:33.1 & 123.9 & 2486.8008 &   17.864 &  16.464 &  14.294 &
H-V13, A-V1292\\ 
cep002 & 19:44:56.82 & -14:53:15.7 & 65.32 & 2566.5870 &   17.677 &  16.428 &  14.492 & H-V7\\ 
cep003 & 19:44:55.23 & -14:47:12.7 & 37.461 & 2541.6318 &   18.688 &  17.423 &  15.462 &
H-V2, A-V3873\\ 
cep004 & 19:44:56.76 & -14:51:18.5 & 34.663 & 2565.6012 &   18.865 &  17.578 &  15.583 & K-V28\\ 
cep005 & 19:45:25.05 & -14:58:44.8 & 32.419 & 2492.7951 &   18.840 &  17.664 &  15.841 & \\ 
cep006 & 19:45:15.81 & -14:55:24.2 & 31.868 & 2577.5338 &   18.560 &  17.397 &  15.594
& K-V29 \\ 
cep007 & 19:44:59.72 & -14:47:10.8 & 30.512 & 2489.7943 &   19.014 &  17.736 &  15.755 &
H-V1, V3736\\ 
cep008 & 19:44:53.05 & -14:46:56.6 & 29.211 & 2502.7439 &   19.219 &  17.931 &  15.935
& H-V3, A-V2374 \\ 
cep009 & 19:45:43.30 & -14:45:43.1 & 21.120 & 2529.7086 &   19.416 &  18.276 &  16.509 & \\ 
cep010 & 19:45:02.22 & -14:53:44.2 & 19.960 & 2599.5192 &   19.454 &  18.249 &  16.381 &H-V6 \\ 
cep011 & 19:44:52.42 & -14:47:36.4 & 19.887 & 2770.8827 &   19.878 &  18.523 &  16.423 & A-V1286\\ 
cep012 & 19:45:08.26 & -14:53:49.9 & 19.602 & 2566.5870 &   19.759 &  18.457 &  16.439 & \\ 
cep013 & 19:44:47.18 & -14:49:21.9 & 19.295 & 2489.7943 &   19.715 &  18.583 &  16.828
& K-V17 \\ 
cep014 & 19:45:06.44 & -14:51:04.8 & 18.339 & 2500.7632 &   19.562 &  18.352 &  16.476 & K-V21\\ 
cep015 & 19:44:56.56 & -14:50:34.3 & 17.344 & 2502.7439 &   19.500 &  18.371 &  16.621 &H-V4 \\ 
cep016 & 19:44:51.32 & -14:54:19.1 & 16.960 & 2578.5411 &   19.872 &  18.699 &  16.881 & \\ 
cep017 & 19:44:53.09 & -14:48:39.8 & 16.855 & 2564.6135 &   20.117 &  18.780 &  16.708
&H-V9, A-V0734 \\ 
cep018 & 19:45:02.54 & -14:51:06.8 & 13.872 & 2548.6332 &   19.890 &  18.672 &  16.784 &H-V5 \\ 
cep019 & 19:44:50.35 & -14:49:13.6 & 11.164 & 2511.7407 &   19.978 &  18.823 &  17.033 & \\ 
cep020 & 19:45:23.48 & -14:46:19.0 & 10.925 & 2575.5684 &   19.525 &  18.677 &  17.363 & \\ 
cep021 & 19:45:46.63 & -14:56:42.4 & 10.783 & 2490.7696 &   19.865 &  18.899 &  17.402 & \\ 
cep022 & 19:44:50.16 & -14:52:29.0 & 10.277 & 2578.5411 &   20.544 &  19.382 &  17.581 & \\ 
cep023 & 19:45:04.55 & -14:55:08.5 &  9.5731 & 2492.7951 &   20.205 &  19.124 &  17.448 & \\ 
cep024 & 19:44:50.94 & -14:48:37.3 &  9.3664 & 2574.5653 &   20.453 &  19.256 &  17.401
& K-V25, A-V0729 \\ 
cep025 & 19:45:00.46 & -14:47:04.8 &  8.9367 & 2464.8644 &   20.933 &  19.236 &  16.606 & A-V3738\\ 
cep026 & 19:45:04.46 & -14:52:53.5 &  8.4670 & 2518.7165 &   20.249 &  19.131 &  17.398 & \\ 
cep027 & 19:45:04.64 & -14:53:40.6 &  8.3912 & 2497.7874 &   20.561 &  19.387 &  17.567 & \\ 
cep028 & 19:45:05.65 & -14:53:21.0 &  7.2085 & 2554.6080 &   19.920 &  18.529 &  16.373 & \\ 
cep029 & 19:45:02.93 & -14:51:57.3 &  7.1054 & 2471.8358 &   20.918 &  19.749 &  17.937 & \\ 
cep030 & 19:44:52.00 & -14:45:52.1 &  6.8842 & 2521.7141 &   20.965 &  19.771 &  17.920 & \\ 
cep031 & 19:44:51.12 & -14:47:20.6 &  6.8665 & 2489.7943 &   20.819 &  19.717 &  18.009 & A-A1433\\ 
cep032 & 19:45:27.86 & -14:53:52.4 &  6.7650 & 2517.7361 &   20.727 &  19.743 &  18.218 & \\ 
cep033 & 19:45:01.63 & -14:49:21.8 &  6.7356 & 2496.7490 &   20.959 &  19.675 &  17.685 & A-V0372\\ 
\enddata
\end{deluxetable}

\setcounter{table}{0}
\begin{deluxetable}{c c c c c c c c c}
\tablecaption{Cepheids in NGC 6822 - continued}
\tablehead{
\colhead{ID} & \colhead{RA} & \colhead{DEC}  & \colhead{P} & \colhead{
${\rm
T}_{0}$} &
\colhead{$<V>$} & \colhead{$<I>$} & \colhead{$<W_{\rm I}>$} &
\colhead{Remarks} \\
 & \colhead{(J2000)} & \colhead{(J2000)}  &
\colhead{ [days]} & \colhead{-2450000} &
\colhead{[mag]} & \colhead{[mag]} & \colhead{[mag]} &
}
\startdata

cep034 & 19:45:08.76 & -14:49:10.7 &  6.5391 & 2460.8596 &   20.739 &  19.541 &  17.684 & \\ 
cep035 & 19:45:01.58 & -14:43:10.1 &  6.2965 & 2770.8827 &   20.754 &  19.665 &  17.977 & \\ 
cep036 & 19:45:19.50 & -14:45:58.4 &  6.1457 & 2463.8594 &   20.568 &  19.625 &  18.163 & \\ 
cep037 & 19:44:51.06 & -14:46:06.0 &  6.1387 & 2545.6034 &   21.074 &  20.030 &  18.412 & \\ 
cep038 & 19:44:30.71 & -14:51:22.1 &  6.0637 & 2548.6332 &   20.904 &  19.817 &  18.132 & \\ 
cep039 & 19:45:23.88 & -14:46:33.3 &  6.0011 & 2458.8719 &   20.734 &  19.691 &  18.074 & \\ 
cep040 & 19:45:36.21 & -14:54:56.5 &  5.9793 & 2552.6243 &   21.236 &  19.424 &  16.615 & \\ 
cep041 & 19:45:04.66 & -14:49:39.5 &  5.9760 & 2516.7308 &   20.498 &  19.543 &  18.063 & A-V0271\\ 
cep042 & 19:44:12.78 & -14:41:08.3 &  5.7100 & 2464.8644 &   20.811 &  19.857 &  18.378 & \\ 
cep043 & 19:45:00.93 & -14:47:50.1 &  5.6450 & 2463.8594 &   20.913 &  19.835 &  18.164 & \\ 
cep044 & 19:44:33.77 & -14:39:36.1 &  5.5905 & 2557.6145 &   20.919 &  19.862 &  18.224 & \\ 
cep045 & 19:44:44.05 & -14:51:26.8 &  5.5621 & 2463.8594 &   21.031 &  19.941 &  18.251 & \\ 
cep046 & 19:44:54.16 & -14:49:47.9 &  5.5518 & 2787.8786 &   20.992 &  19.953 &  18.343 & \\ 
cep047 & 19:45:23.68 & -14:49:03.5 &  5.3611 & 2584.5470 &   20.787 &  19.850 &  18.398 & \\ 
cep048 & 19:44:49.45 & -14:46:57.0 &  5.1764 & 2591.5357 &   21.523 &  20.333 &  18.488 & \\ 
cep049 & 19:45:21.67 & -14:53:56.9 &  5.1701 & 2548.6332 &   21.022 &  20.252 &  19.059 & \\ 
cep050 & 19:45:08.19 & -14:41:27.5 &  5.1285 & 2498.7581 &   21.146 &  20.214 &  18.769 & \\ 
cep051 & 19:45:01.67 & -14:51:27.8 &  5.1090 & 2579.5495 &   20.686 &  19.898 &  18.677 & \\ 
cep052 & 19:45:08.45 & -14:46:43.1 &  5.0970 & 2488.7265 &   20.998 &  20.046 &  18.570 & \\ 
cep053 & 19:44:25.34 & -14:43:39.4 &  5.0773 & 2518.7165 &   21.335 &  20.234 &  18.527 & \\ 
cep054 & 19:45:12.08 & -14:44:43.5 &  5.0560 & 2546.6257 &   21.056 &  20.162 &  18.776 & \\ 
cep055 & 19:45:11.37 & -14:49:53.1 &  4.8955 & 2496.7490 &   21.086 &  20.021 &  18.370 & \\ 
cep056 & 19:44:56.64 & -14:45:55.8 &  4.8822 & 2529.7086 &   21.297 &  20.397 &  19.002 & \\ 
cep057 & 19:45:08.01 & -14:54:12.2 &  4.8305 & 2562.5883 &   20.880 &  19.992 &  18.616 & \\ 
cep058 & 19:44:53.84 & -14:49:06.7 &  4.6860 & 2486.8008 &   21.542 &  20.180 &  18.069 & \\ 
cep059 & 19:44:54.24 & -14:43:18.0 &  4.6345 & 2521.7141 &   21.050 &  20.057 &  18.518 & \\ 
cep060 & 19:44:37.93 & -14:48:31.0 &  4.6046 & 2520.7108 &   21.196 &  20.100 &  18.401 & A-V0771\\ 
cep061 & 19:44:59.35 & -14:47:45.3 &  4.5480 & 2458.8719 &   21.275 &  20.031 &  18.103 & A-V3186\\ 
cep062 & 19:44:59.64 & -14:44:34.2 &  4.5264 & 2593.5212 &   21.094 &  20.053 &  18.439 & \\ 
cep063 & 19:44:50.75 & -14:46:20.1 &  4.4495 & 2526.6949 &   21.005 &  19.701 &  17.680 & \\ 
cep064 & 19:44:54.70 & -14:51:10.9 &  4.3602 & 2555.6347 &   21.121 &  19.973 &  18.194 & \\ 
cep065 & 19:44:19.00 & -14:40:48.8 &  4.2612 & 2574.5653 &   21.067 &  20.066 &  18.514 & \\ 
cep066 & 19:44:10.57 & -14:45:32.8 &  4.2330 & 2496.7490 &   21.193 &  20.270 &  18.839 & \\ 
cep067 & 19:45:08.98 & -14:54:13.8 &  4.2327 & 2488.7265 &   21.205 &  20.147 &  18.507 & \\ 
\enddata
\end{deluxetable}

\setcounter{table}{0}
\begin{deluxetable}{c c c c c c c c c}
\tablecaption{Cepheids in NGC 6822 - continued}
\tablehead{
\colhead{ID} & \colhead{RA} & \colhead{DEC}  & \colhead{P} & \colhead{
${\rm
T}_{0}$} &
\colhead{$<V>$} & \colhead{$<I>$} & \colhead{$<W_{\rm I}>$} &
\colhead{Remarks} \\
 & \colhead{(J2000)} & \colhead{(J2000)}  &
\colhead{ [days]} & \colhead{-2450000} &
\colhead{[mag]} & \colhead{[mag]} & \colhead{[mag]} &
}
\startdata

cep068 & 19:44:51.23 & -14:50:47.8 &  4.2294 & 2556.6132 &   21.137 &  20.214 &  18.783 & \\ 
cep069 & 19:45:05.54 & -14:52:20.7 &  4.2188 & 2555.6347 &   21.405 &  20.310 &  18.613 & \\ 
cep070 & 19:44:50.69 & -14:50:02.5 &  4.1123 & 2556.6132 &   21.375 &  20.364 &  18.797 & \\ 
cep071 & 19:45:49.90 & -14:53:41.9 &  4.0912 & 2493.7883 &   21.412 &  20.359 &  18.727 & \\ 
cep072 & 19:44:55.25 & -14:49:22.2 &  4.0195 & 2552.6243 &   21.334 &  20.459 &  19.103 & A-V0410\\ 
cep073 & 19:44:49.90 & -14:51:42.1 &  3.9879 & 2556.6132 &   21.523 &  20.634 &  19.256 & \\ 
cep074 & 19:45:37.87 & -14:54:11.1 &  3.9362 & 2487.7294 &   21.245 &  20.365 &  19.001 & \\ 
cep075 & 19:44:47.97 & -14:44:28.6 &  3.8970 & 2517.7361 &   21.427 &  20.416 &  18.849 & \\ 
cep076 & 19:44:57.33 & -14:49:10.8 &  3.7880 & 2490.7696 &   20.007 &  19.255 &  18.089 & A-V0444\\ 
cep077 & 19:44:59.98 & -14:51:27.7 &  3.7751 & 2500.7632 &   21.119 &  19.919 &  18.059 & \\ 
cep078 & 19:44:59.37 & -14:50:19.1 &  3.7180 & 2547.6173 &   21.072 &  20.028 &  18.410 & \\ 
cep079 & 19:44:46.87 & -14:50:26.6 &  3.7079 & 2526.6949 &   21.080 &  19.531 &  17.130 & \\ 
cep080 & 19:45:02.95 & -14:43:30.9 &  3.6800 & 2488.7265 &   21.822 &  20.560 &  18.604 & \\ 
cep081 & 19:44:57.24 & -14:49:51.4 &  3.6479 & 2526.6949 &   21.635 &  20.587 &  18.963 & A-V0171\\ 
cep082 & 19:44:36.14 & -14:43:35.0 &  3.6190 & 2490.7696 &   21.427 &  20.494 &  19.048 & \\ 
cep083 & 19:44:49.96 & -14:54:29.5 &  3.5845 & 2460.8596 &   21.313 &  20.059 &  18.115 & \\ 
cep084 & 19:44:09.68 & -14:41:23.1 &  3.5468 & 2770.8827 &   21.133 &  20.431 &  19.343 & \\ 
cep085 & 19:45:15.76 & -14:44:08.3 &  3.5078 & 2463.8594 &   21.278 &  20.402 &  19.044 & \\ 
cep086 & 19:44:53.49 & -14:50:21.4 &  3.4584 & 2546.6257 &   21.633 &  20.533 &  18.828 & \\ 
cep087 & 19:44:51.80 & -14:53:00.9 &  3.3700 & 2546.6257 &   21.501 &  20.834 &  19.800 & \\ 
cep088 & 19:44:11.68 & -14:49:37.0 &  3.3659 & 2563.5980 &   21.569 &  21.790 &  22.133 & \\ 
cep089 & 19:45:14.27 & -14:53:14.9 &  3.3032 & 2520.7108 &   21.544 &  20.449 &  18.752 & \\ 
cep090 & 19:45:47.48 & -14:54:24.7 &  3.2916 & 2770.8827 &   21.430 &  20.504 &  19.069 & \\ 
cep091 & 19:44:36.19 & -14:45:07.1 &  3.2780 & 2550.6265 &   21.511 &  20.501 &  18.935 & \\ 
cep092 & 19:44:02.24 & -14:38:02.0 &  3.2273 & 2498.7581 &   21.178 &  99.999 & 99.999 & \\ 
cep093 & 19:45:12.01 & -14:53:36.3 &  3.2069 & 2492.7951 &   21.591 &  20.712 &  19.350 & \\ 
cep094 & 19:44:58.35 & -14:50:31.3 &  3.1756 & 2546.6257 &   21.525 &  20.413 &  18.689 & \\ 
cep095 & 19:44:45.50 & -14:49:29.7 &  3.1093 & 2576.5305 &   21.637 &  20.824 &  19.564 & \\ 
cep096 & 19:45:09.50 & -14:44:21.9 &  2.8979 & 2496.7490 &   21.608 &  20.597 &  19.030 & \\ 
cep097 & 19:44:57.74 & -14:52:01.4 &  2.8486 & 2502.7439 &   21.139 &  20.226 &  18.811 & \\ 
cep098 & 19:44:43.98 & -14:51:53.1 &  2.6760 & 2559.5401 &   21.913 &  21.141 &  19.944 & \\ 
cep099 & 19:44:58.34 & -14:53:18.2 &  2.6684 & 2547.6173 &   21.944 &  20.797 &  19.019 & \\ 
cep100 & 19:44:49.56 & -14:45:10.1 &  2.5950 & 2547.6173 &   21.077 &  20.229 &  18.915 & \\ 
cep101 & 19:45:06.90 & -14:53:21.4 &  2.5937 & 2578.5411 &   21.391 &  19.744 &  17.191 & \\ 
\enddata
\end{deluxetable}

\setcounter{table}{0}
\begin{deluxetable}{c c c c c c c c c}
\tablecaption{Cepheids in NGC 6822 - continued}
\tablehead{
\colhead{ID} & \colhead{RA} & \colhead{DEC}  & \colhead{P} & \colhead{
${\rm
T}_{0}$} &
\colhead{$<V>$} & \colhead{$<I>$} & \colhead{$<W_{\rm I}>$} &
\colhead{Remarks} \\
 & \colhead{(J2000)} & \colhead{(J2000)}  &
\colhead{ [days]} & \colhead{-2450000} &
\colhead{[mag]} & \colhead{[mag]} & \colhead{[mag]} &
}
\startdata
cep102 & 19:45:21.13 & -14:48:33.2 &  2.5839 & 2577.5338 &   21.554 &  20.343 &  18.466 & \\ 
cep103 & 19:44:54.87 & -14:50:26.2 &  2.5390 & 2517.7361 &   21.849 &  20.481 &  18.361 & \\ 
cep104 & 19:44:34.06 & -14:38:21.5 &  2.4540 & 2577.5338 &   21.920 &  99.999 & 99.999 & \\ 
cep105 & 19:44:21.82 & -14:41:49.7 &  2.4536 & 2546.6257 &   21.760 &  20.987 &  19.789 & \\ 
cep106 & 19:45:51.25 & -14:49:09.2 &  2.3725 & 2593.5212 &   19.162 &  17.719 &  15.482 & \\ 
cep107 & 19:45:08.83 & -14:46:10.1 &  2.3571 & 2500.7632 &   21.672 &  20.724 &  19.255 & \\ 
cep108 & 19:45:41.90 & -14:36:17.2 &  2.3124 & 2557.6145 &   21.887 &  21.163 &  20.041 & \\ 
cep109 & 19:45:10.28 & -14:53:02.0 &  2.1755 & 2471.8358 &   21.993 &  21.372 &  20.409 & \\ 
cep110 & 19:45:51.25 & -14:57:43.7 &  2.0915 & 2554.6080 &   22.053 &  20.838 &  18.955 & \\ 
cep111 & 19:45:35.61 & -14:49:11.2 &  2.0772 & 2555.6347 &   21.162 &  20.410 &  19.244 & \\ 
cep112 & 19:45:39.71 & -14:38:32.8 &  2.0740 & 2559.5401 &   22.253 &  21.395 &  20.065 & \\ 
cep113 & 19:45:09.73 & -14:55:09.6 &  2.0697 & 2487.7294 &   20.942 &  19.860 &  18.183 & \\ 
cep114 & 19:45:33.97 & -14:54:53.7 &  2.0645 & 2559.5401 &   21.255 &  20.518 &  19.376 & \\ 
cep115 & 19:44:38.11 & -14:52:41.1 &  1.8683 & 2579.5495 &   22.262 &  21.432 &  20.145 & \\ 
cep116 & 19:45:05.95 & -14:51:23.0 &  1.7092 & 2486.8008 &   21.547 &  20.171 &  18.038 & \\ 
\enddata
\end{deluxetable}

\begin{deluxetable}{ccccc}
\tablecaption{Individual V and I Observations}
\tablehead{
\colhead{object}  & \colhead{filter} &
\colhead{HJD-2450000}  & \colhead{mag}  & \colhead{$\sigma_{mag}$}\\
}
\startdata
cep001 & V & 2458.87193 &  18.044  &   0.014 \\ 
cep001 & V & 2460.85964 &  18.059  &   0.018 \\ 
cep001 & V & 2463.85945 &  18.124  &   0.013 \\ 
cep001 & V & 2464.86437 &  18.143  &   0.013 \\ 
cep001 & V & 2465.85531 &  18.149  &   0.014 \\ 
cep001 & V & 2468.85143 &  18.207  &   0.014 \\ 
cep001 & V & 2470.84386 &  18.237  &   0.015 \\ 
cep001 & V & 2471.83581 &  18.231  &   0.014 \\ 
cep001 & V & 2482.70757 &  18.327  &   0.032 \\ 
cep001 & V & 2483.78027 &  18.358  &   0.024 \\ 
cep001 & V & 2486.80078 &  18.370  &   0.023 \\ 
cep001 & V & 2487.72942 &  18.367  &   0.021 \\ 
cep001 & V & 2488.72654 &  18.352  &   0.019 \\ 
cep001 & V & 2489.79432 &  18.366  &   0.012 \\ 
cep001 & V & 2490.76962 &  18.337  &   0.020 \\ 
cep001 & V & 2492.79510 &  18.313  &   0.020 \\ 
cep001 & V & 2493.78829 &  18.344  &   0.016 \\ 
cep001 & V & 2495.79027 &  18.258  &   0.016 \\ 
cep001 & V & 2496.74902 &  18.248  &   0.013 \\ 
cep001 & V & 2497.78736 &  18.231  &   0.015 \\ 
\enddata
\tablecomments{The complete version of this table is in the electronic
edition of the Journal.  The printed edition contains only
the the first 20 measurements in V band for the Cepheid variable cep001.}

\end{deluxetable}

\begin{figure}[htb] 
\vspace*{18cm}
\includegraphics{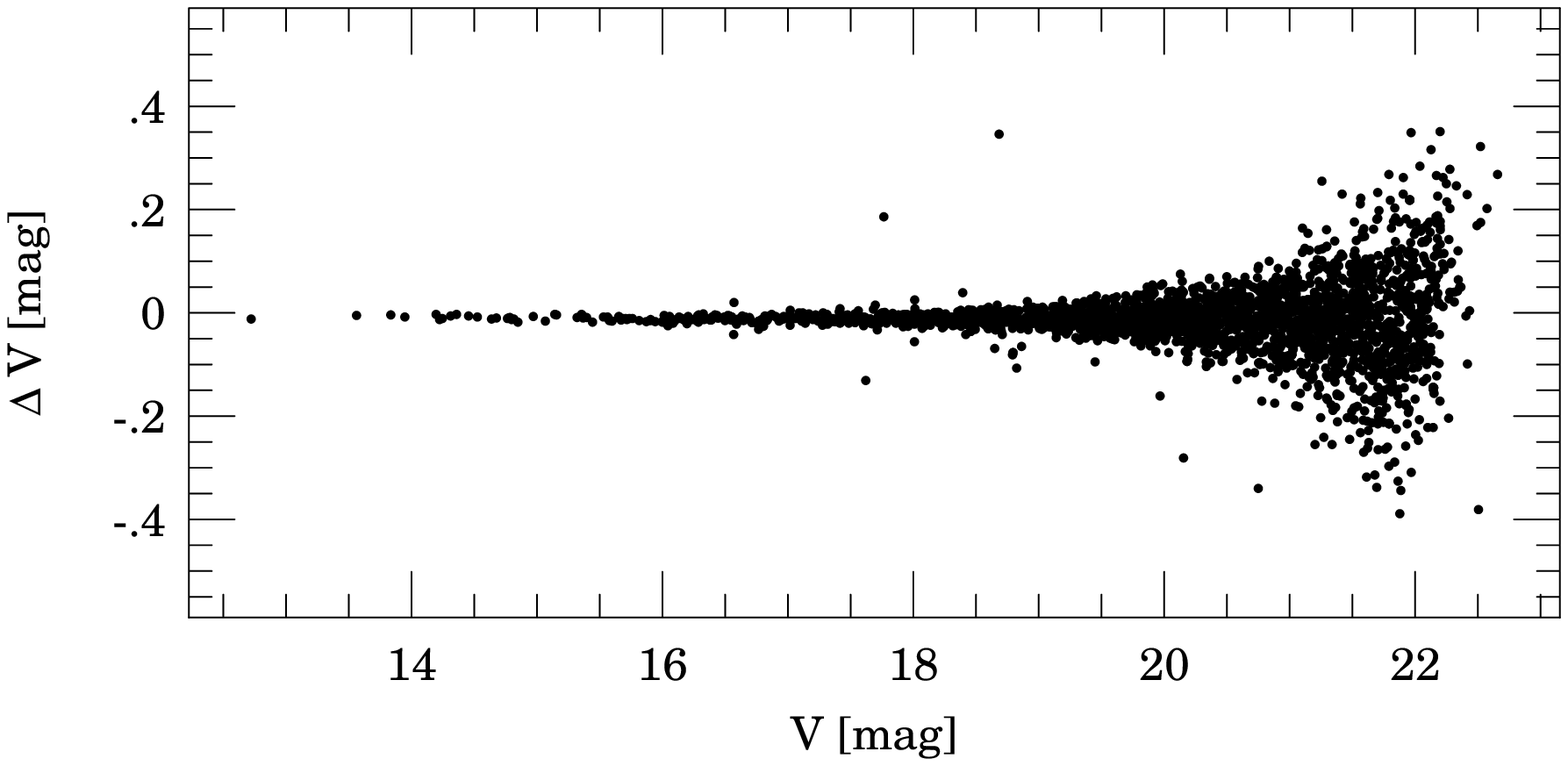} 
\includegraphics{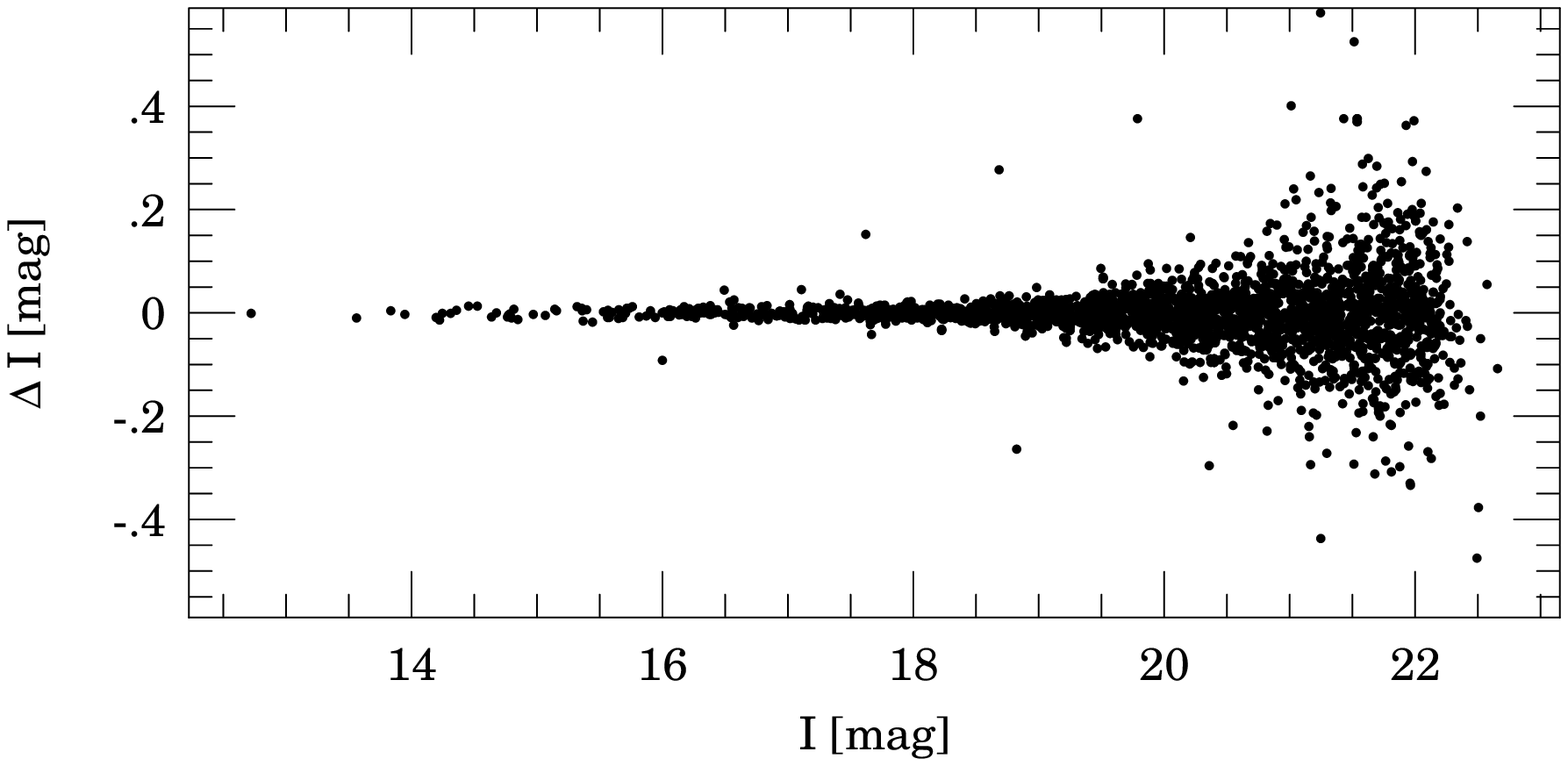}
\caption{Comparison of V and I band magnitudes of stars in NGC 6822 
observed and calibrated on two different nights on one of the chips of the mosaic camera. 
An excellent (always better than 0.01 mag) internal agreement is demonstrated.}
\end{figure}  

\begin{figure}[htb]
\vspace*{18cm}
\includegraphics{pietrzynski.fig2a.ps}
\includegraphics{pietrzynski.fig2b.ps}
\caption{Comparison of OGLE II, and our new mosaic photometry in the SMC fields (see text),
for one chip.  Additional scatter (up to  0.05 mag) is clearly seen, and 
most probably it is caused by variations of the zero points over the
mosaic field.  }
\end{figure}

\begin{figure}[htb]
\vspace*{18cm}
\includegraphics{pietrzynski.fig3a.ps}
\includegraphics{pietrzynski.fig3b.ps}
\caption{Comparison of OGLE II, and our new photometry in the SMC fields, 
after correcting our data for variations of the zero point over the mosaic field- see text 
for details. This plot demonstrates that our correction maps are useful 
to correct the variations of the zero point over the mosaic down to 
about 0.03 mag.}
\end{figure}

\begin{figure}[htb]
\vspace*{18cm}
\includegraphics{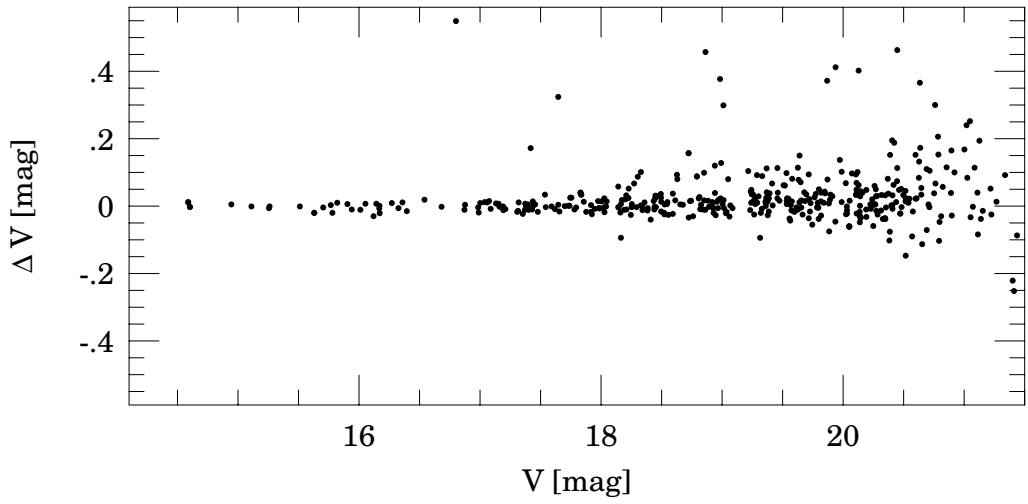}
\caption{Comparison of our NGC 6822 V-band photometry with the photometric data published 
by  Bianchi et al. (2001).}
\end{figure}

\begin{figure}[htb]
\vspace*{18cm}
\includegraphics{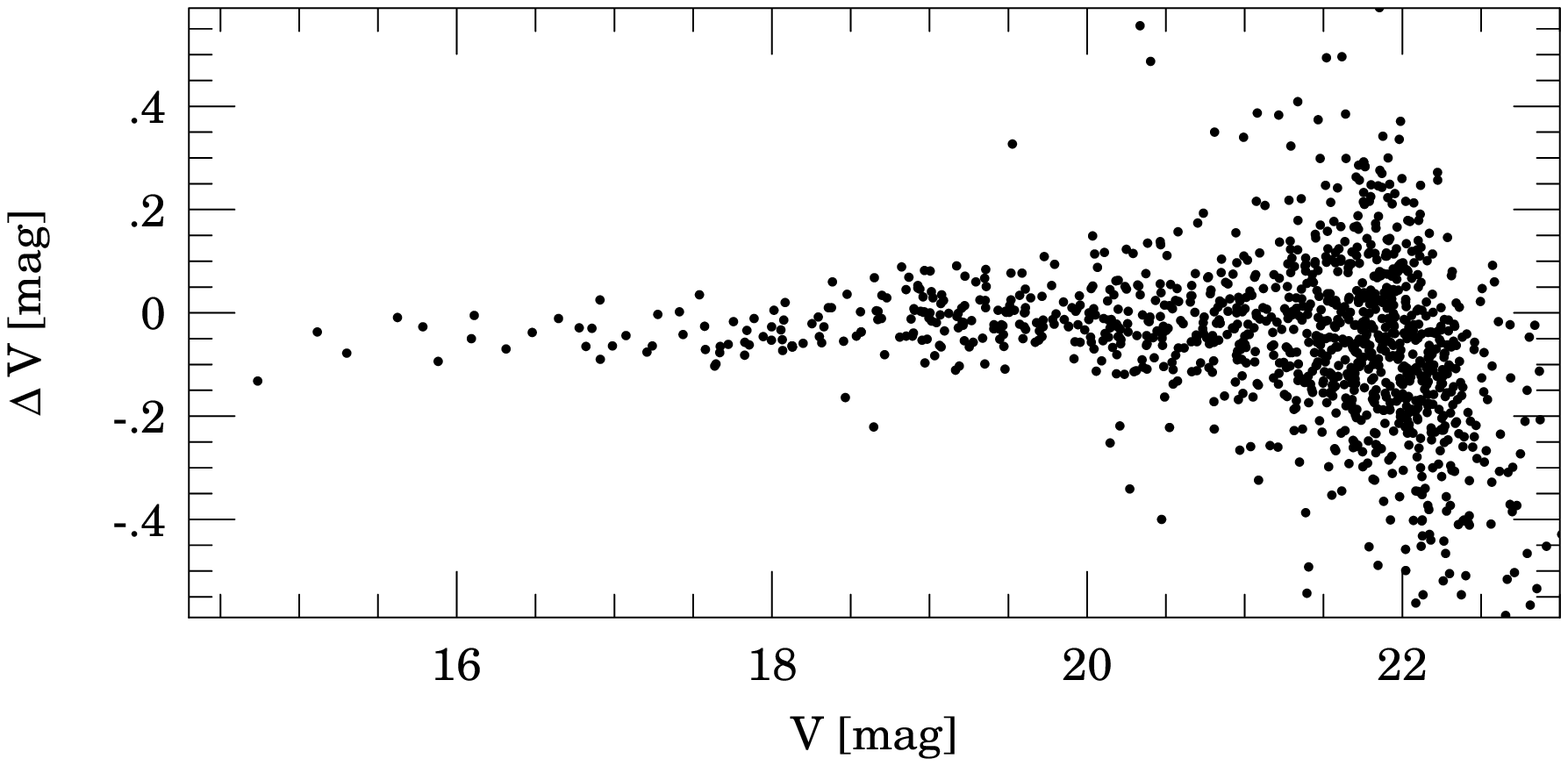}
\includegraphics{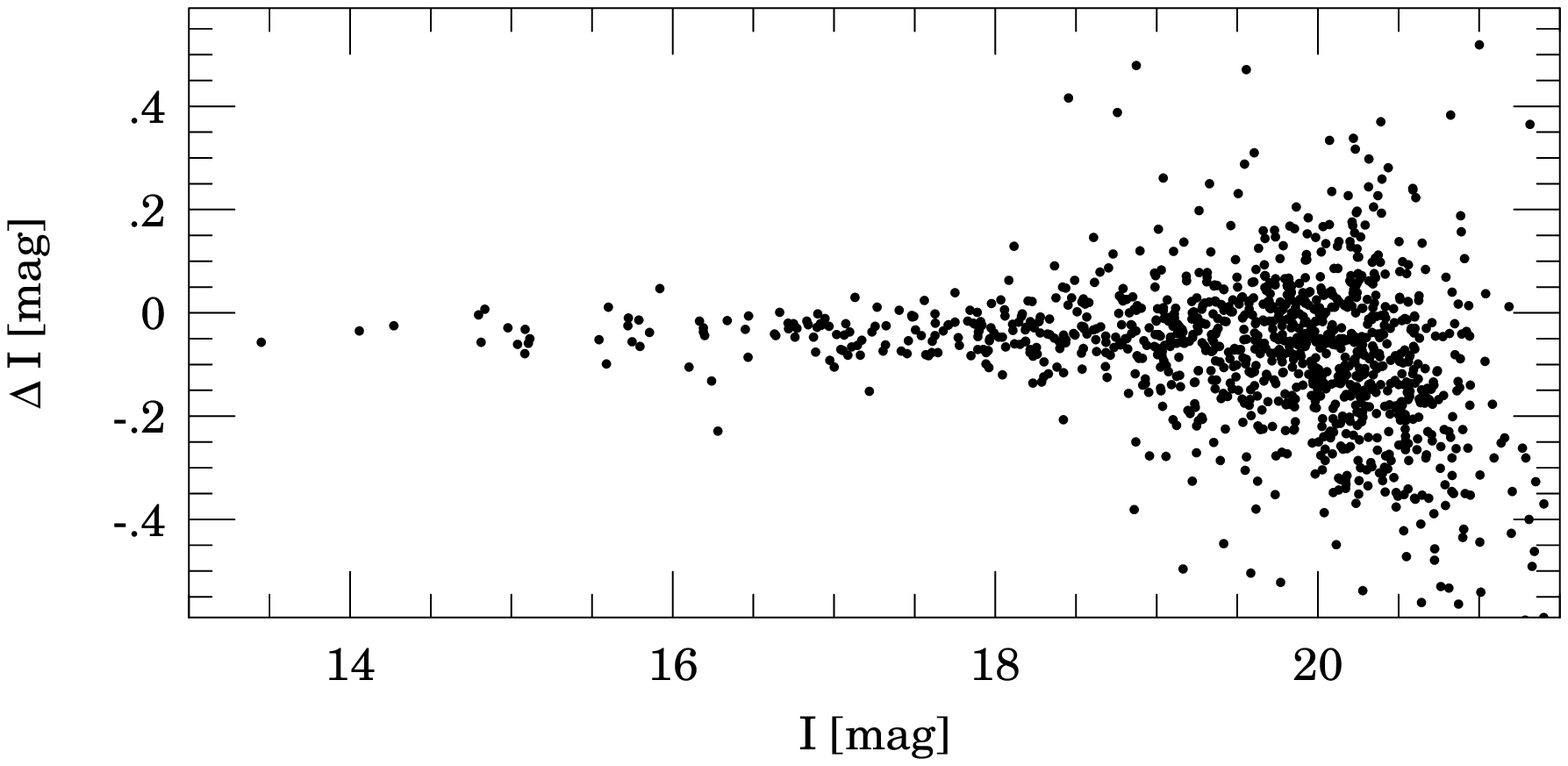}
\caption{Comparison of our NGC 6822 V- and I-band photometry with the data published
by  Gallart et al. (1996).}
\end{figure}

\begin{figure}[htb]
\vspace*{22cm}
\includegraphics{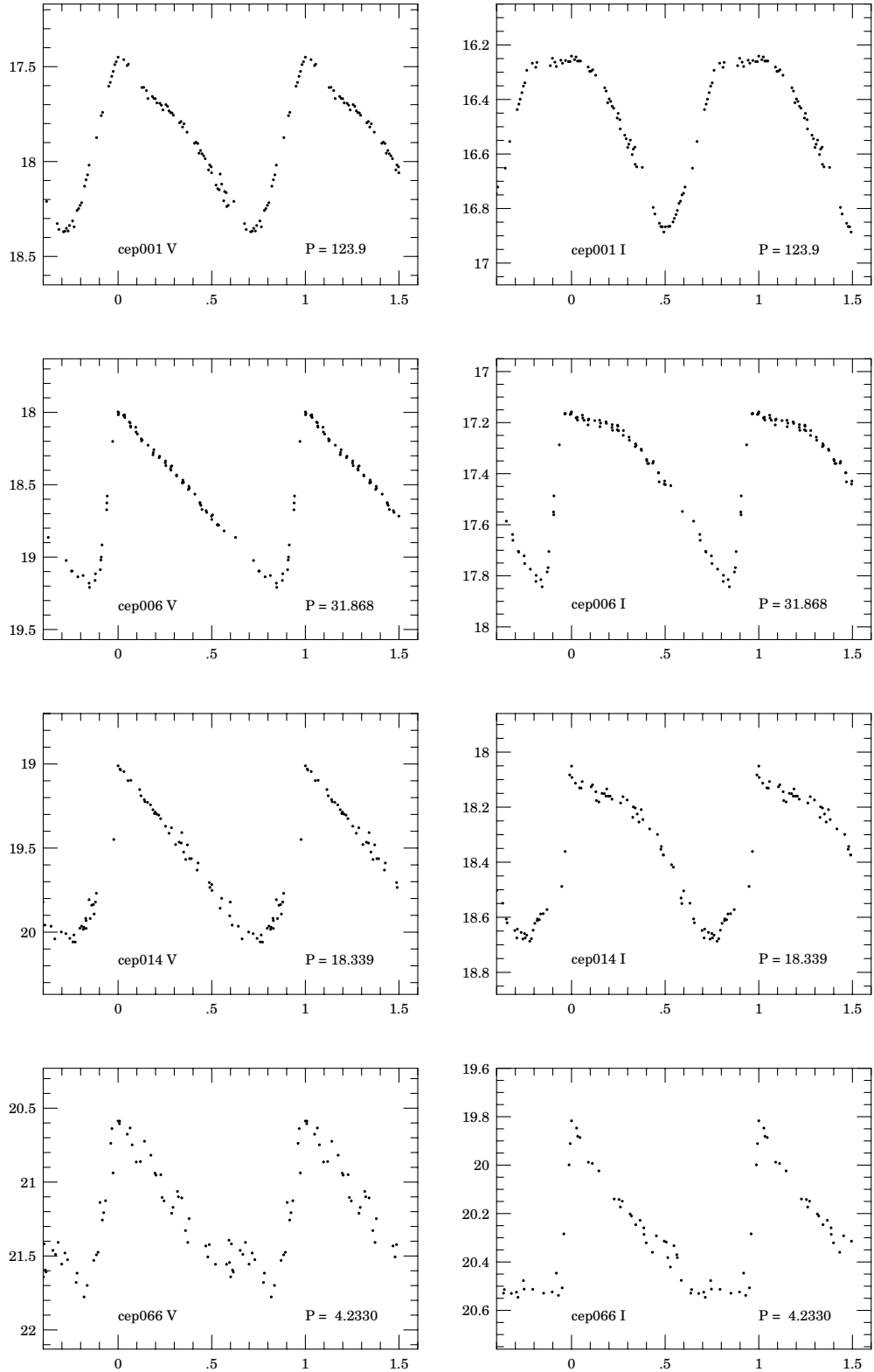}
\caption{Phased V- and I-band light curves for some Cepheids of different
periods in our NGC 6822 catalog. These light curves are representative for the
light curves of other Cepheid variables of similar periods.
}
\end{figure}

\begin{figure}[htb]
\vspace*{15cm}
\includegraphics{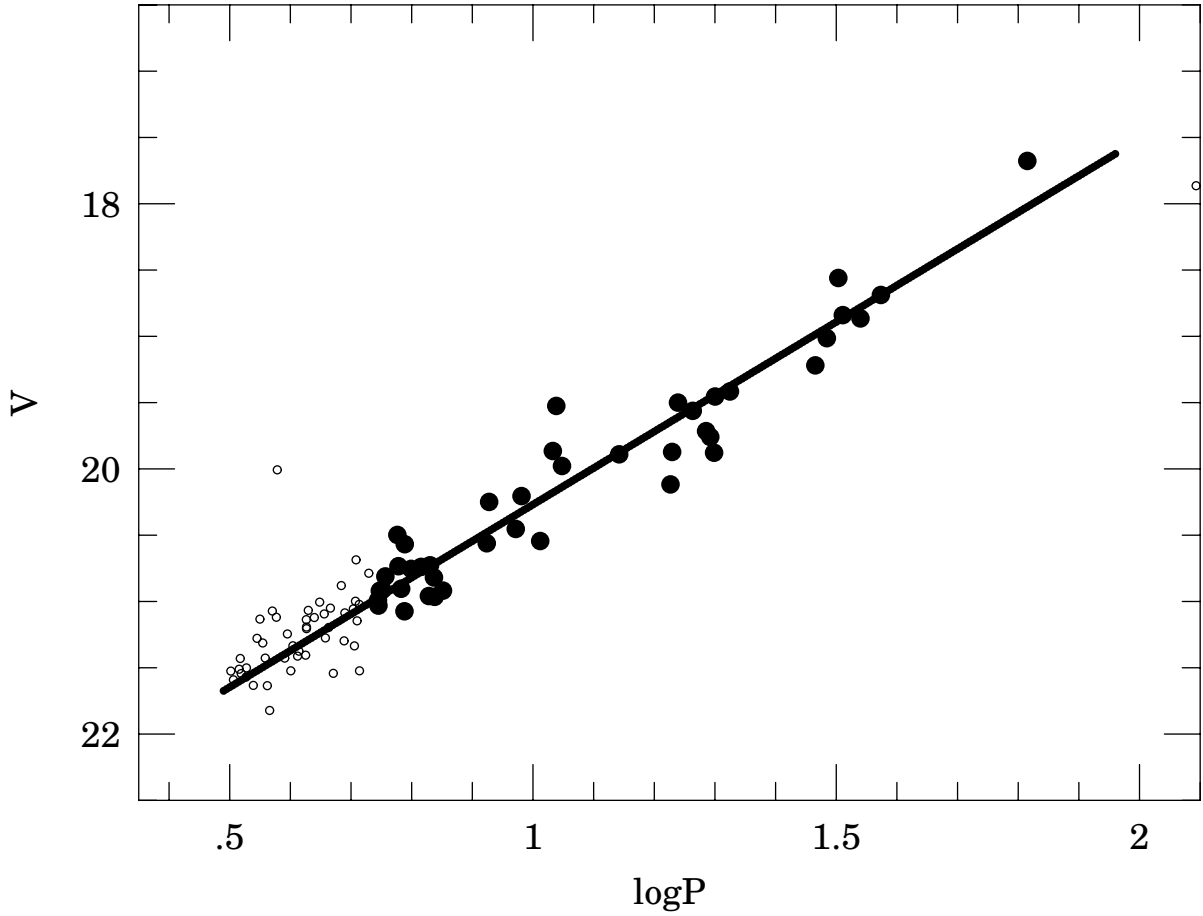}
\caption{The period-luminosity relation for NGC 6822 Cepheids in the V 
band. Open circles denote variables not used in the distance determination
because some of these stars might be overtone pulsators.
The slope of the relation was adopted from the LMC Cepheids (OGLE II).
}
\end{figure}

\begin{figure}[htb]
\vspace*{15cm}
\includegraphics{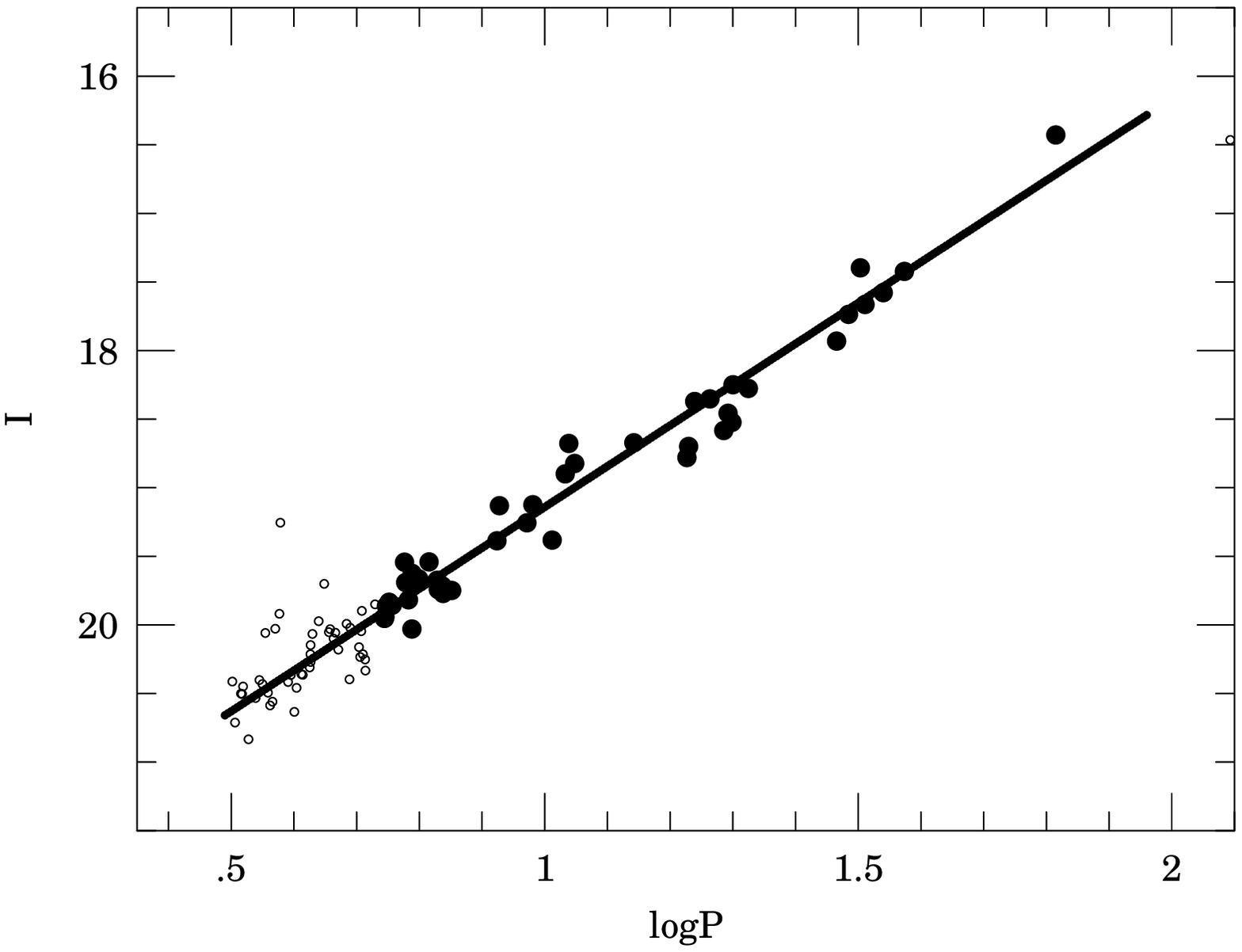}
\caption{Same as Fig. 6, for the I band.
}
\end{figure}

\begin{figure}[htb]
\vspace*{15cm}
\includegraphics{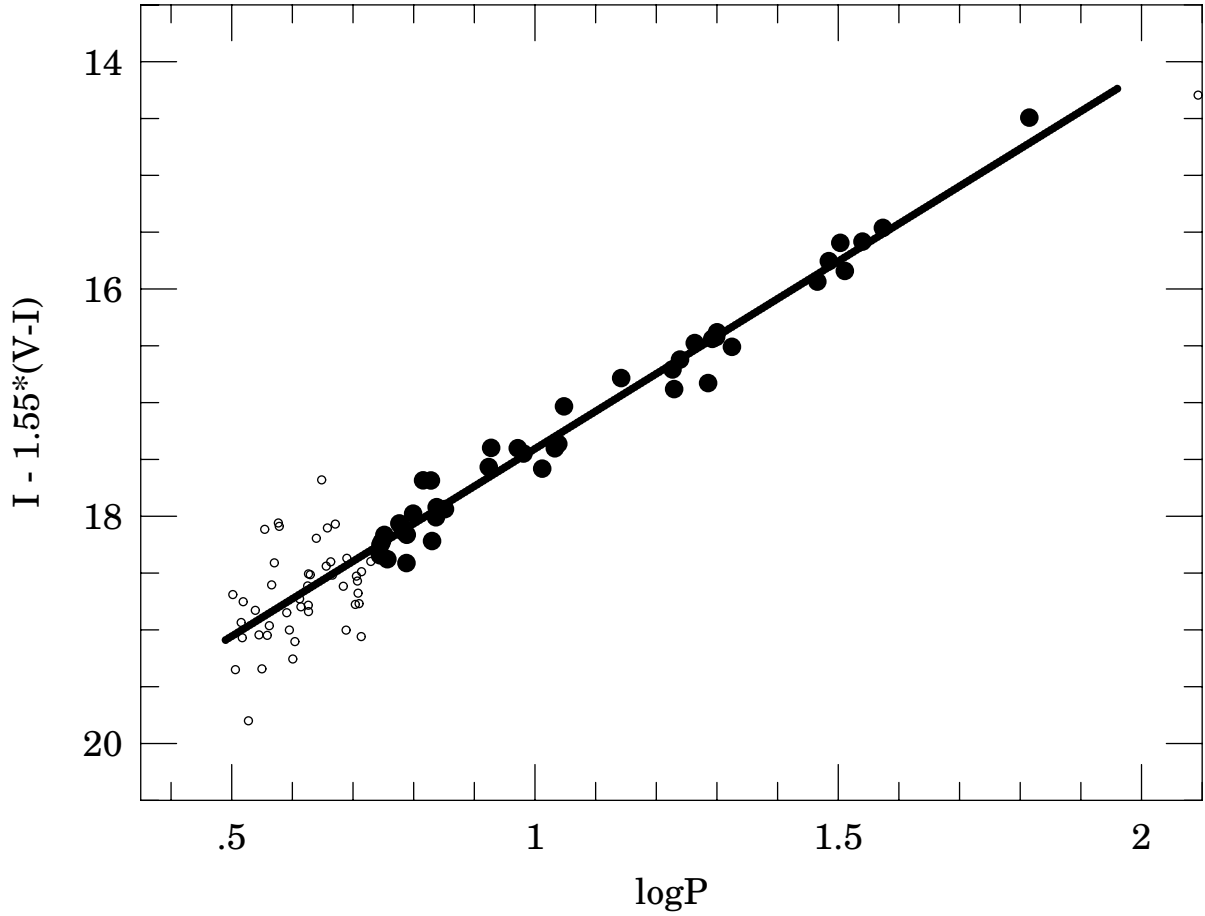}
\caption{Same as Fig. 6, for the reddening-independent (V-I) Wesenheit 
magnitudes. We note that using all the stars (open circles included)
in our distance determination to NGC 6822, the result changes in a
negligible way.
}
\end{figure}

\begin{figure}[htb]
\vspace*{8 cm}
\includegraphics{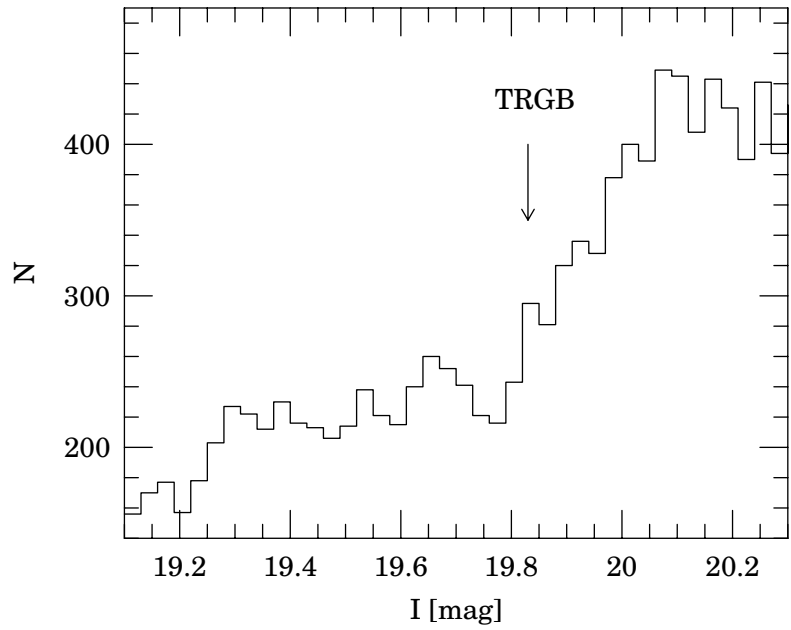}
\caption{Luminosity function of the red giant branch in NGC 6822 
constructed from our mosaic data using bins of 0.05 mag. The arrow shows 
the position of the TRGB magnitude.
}
\end{figure}

\begin{figure}[htb]
\vspace*{10 cm}
\includegraphics{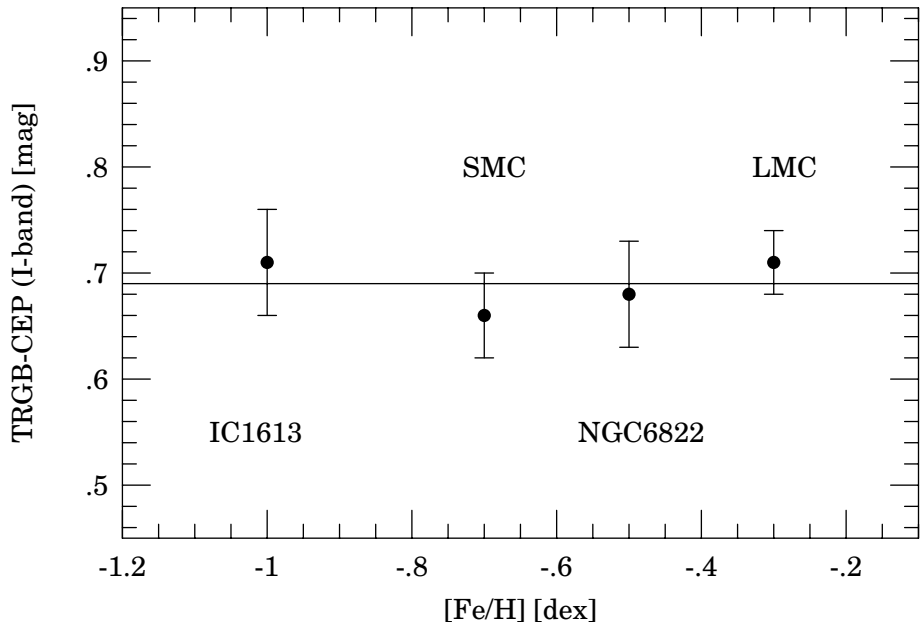}
\caption{Comparison of the I-band brightness of Cepheids with a 
period of 10 days, and the I-band magnitude of the tip of the red giant branch
for four galaxies covering a wide range in metallicities, and having
excellent data for the determination of these quantities.
}

\end{figure}

\end{document}